# Cockroaches use diverse strategies to self-right on the ground

Chen Li[1,2*], Toni Wöhrl[3], Han K. Lam[2], Robert J. Full[2]

[1]Department of Mechanical Engineering, Johns Hopkins University

[2]Department of Integrative Biology, University of California, Berkeley

[3]Institute of Sports Science, Friedrich-Schiller-Universität Jena

*

## SUMMARY STATEMENT

Comparative study of cockroach self-righting reveals performance advantages of using rotational kinetic energy to overcome potential energy barrier and rolling more to lower it, while maintaining diverse strategies.

## ABSTRACT

Terrestrial animals often must self-right from an upside-down orientation on the ground to survive. Here, we compared self-righting strategies of the Madagascar hissing, American, and discoid cockroaches on a challenging flat, rigid, low-friction surface to quantify the mechanical principles. All three species almost always self-righted (97% probability) when given time (30 seconds), frequently self-righted (63%) on the first attempt, and on that attempt did so in one second or less. When successful, two of the three species gained and used pitch and/or roll rotational kinetic energy to overcome potential energy barriers (American 63% of all attempts and discoid 78%). By contrast, the largest, heaviest, wingless cockroach (Madagascar hissing) relied far less on the energy of motion and was the slowest to self-right. Two of the three species used rolling strategies to overcome low potential energy barriers. Successful righting attempts had greater rolling rotation than failed attempts as the center of mass rose to the highest position. Madagascar hissing cockroaches rolled using body deformation (98% of all trials) and the American cockroach relied on leg forces (93%). By contrast, the discoid cockroach overcame higher and a wider range of potential energy barriers with simultaneous pitching and rolling using wings (46% of all trials) and legs (49%) equally to self-right. Our quantification revealed the performance advantages of using rotational kinetic energy to overcome the potential energy barrier and rolling more to lower it, while maintaining diverse strategies for ground-based self-righting.

## INTRODUCTION

Righting oneself from upside down on the ground is a prevalent locomotor transition that many animals must perform to survive. Even on flat, level ground with high friction, legged locomotion can induce large pitch and roll moments (Ting et al., 1994) that can result in overturning. During locomotion in complex terrain with inclinations (Minetti et al., 2002), uneven topology (Chiari et al., 2017; Daley and Biewener, 2006; Sponberg and Full, 2008), low friction (Clark and Higham, 2011), uncertain contact (Spagna et al., 2007), flowable ground (Li et al., 2012), and cluttered obstacles (Li et al., 2015; Li et al., 2017), overturning is even more likely. Other forms of terrestrial locomotion like jumping (Faisal and Matheson, 2001; Libby et al., 2012) and climbing (Jusufi et al., 2008), as well as flying (Faisal and Matheson, 2001) and swimming (Vosatka, 1970), can suffer instability and loss of body control resulting in overturning. Non-locomotor behaviors such as fighting and courtship can also produce overturning (Mann et al., 2006; Willemsen and Hailey, 2003). Under these circumstances, animals must be able to self-right promptly to avoid predation, starvation, and dehydration, as well as to sense, locomote, and reproduce.

Small animals like insects are particularly susceptible to overturning, because they are more sensitive to perturbations resulting from small body inertia (Walter and Carrier, 2002) and terrain irregularities negligible to larger animals (Kaspari and Weiser, 1999). Ground-based self-righting has been studied in many insect species, including beetles (Evans, 1973; Frantsevich, 2004; Frantsevich and Mokrushov, 1980), cockroaches (Camhi, 1977; Delcomyn, 1987; Full et al., 1995; Reingold and Camhi, 1977; Sherman et al., 1977; Zill, 1986), stick insects (Graham, 1979), locusts (Faisal and Matheson, 2001), and springtails (Brackenbury, 1990). Many self-righting strategies have been described (Brackenbury, 1990; Camhi, 1977; Evans, 1973; Faisal and Matheson, 2001; Frantsevich, 2004; Full et al., 1995; Zill, 1986), including: (1) using appendages (legs, wings, tail, and antennae) and head to grasp, pivot, push, or pull, (2) deforming the body, and (3) jumping with elastic energy storage and release. Some insects use multiple strategies and transition among them to self-right (Frantsevich, 2004). In addition, insects can use diverse body rotation including pitching, diagonal rotations (simultaneous pitching and rolling), and rolling (Brackenbury, 1990; Camhi, 1977; Delcomyn, 1987; Evans, 1973; Frantsevich, 2004; Frantsevich and Mokrushov, 1980; Full et al., 1995; Reingold and Camhi, 1977; Sherman et al., 1977; Zill, 1986). Furthermore, neural control and motor patterns of self-righting have been investigated in a variety of insect species (Camhi, 1977; Delcomyn, 1987; Faisal and Matheson, 2001; Frantsevich and Mokrushov, 1980;

Graham, 1979; Reingold and Camhi, 1977; Sherman et al., 1977; Zill, 1986). Although these strategies have been well described, the mechanical principles of ground-based self-righting of small animals remain less understood. Here, we quantify the performance and body rotation of self-righting cockroaches and model the mechanical challenges to gain insight into what governs a small animal's use of various strategies and body rotation.

Previous observations and modeling in turtles have provided insight into the mechanics of how body and appendage morphology affects ground-based self-righting of larger animals (Ashe, 1970; Domokos and Várkonyi, 2008). Ground-based self-righting is the change of body orientation during which the body overcomes gravitational potential energy barriers (Domokos and Várkonyi, 2008). Based on this concept, a planar geometric model explained how shell shape and appendage length together determine whether turtles use active or passive strategies to self-right in the transverse plane (Domokos and Várkonyi, 2008). Turtles primarily rely on passive rotations of unstable shells and/or active, quasi-static pushing of necks and legs to overcome large, primary potential energy barriers. To assist self-righting, turtles also use head and leg bobbing to gain modest amounts of rotational kinetic energy to overcome small, secondary potential energy barriers (Domokos and Várkonyi, 2008). In addition, the dependence of potential energy barriers on body rotation explained why many turtles almost always self-right via body rolling in the transverse plane on level, flat surfaces (Domokos and Várkonyi, 2008; Malashichev, 2016; Rubin et al., 2018; Stancher et al., 2006). Turtles have shells longer in the fore-aft than in the lateral direction, so body pitching overcomes higher potential energy barriers than body rolling does. Because turtles cannot gain sufficient body rotational kinetic energy to overcome the large potential energy barriers required for self-righting using pitching, they roll to self-right.

Here, inspired by these insights, we take the next step in understanding the mechanical principles of ground-based self-righting of small animals. First, we hypothesized that small insects' self-righting strategies can be dynamic, being able to gain and use pitch and/or roll rotational kinetic energy to overcome primary potential energy barriers. Dynamic behavior is plausible because many insects like cockroaches and beetles are capable of rapid locomotion and generating large impulses relative to body weight (Koditschek et al., 2004; Sponberg and Full, 2008; Ting et al., 1994; Zurek and Gilbert, 2014). Second, we hypothesized that, given the diverse three-dimensional body rotations possible (Brackenbury, 1990; Camhi, 1977; Delcomyn, 1987; Evans, 1973; Frantsevich, 2004; Frantsevich and Mokrushov, 1980; Full et al., 1995; Reingold and Camhi, 1977; Sherman et al., 1977; Zill, 1986), insects roll more when they succeed in self-righting than when they fail because increased rolling lowers potential energy barriers.

To test our hypotheses, we studied self-righting on a flat, rigid, low friction surface of three species of cockroaches, the Madagascar hissing cockroach (*Gromphadorhina portentosa*), the American cockroach (*Periplaneta americana*), and the discoid cockroach (*Blaberus discoidalis*), which differ in body size, body shape, leg length, and availability of wings (Fig. 1). The selection of multiple species (Chiari et al., 2017; Domokos and Várkonyi, 2008) from a common super order (Dictyoptera) (Bell et al., 2007) provided access to observing a greater number of strategies and body rotations, but with phylogenetic control that allows comparison. We used high-speed imaging to measure the animals' self-righting performance and body rotation. We used a locomotor transition ethogram analysis to quantify probability distribution of and transitions between self-righting strategies. We developed a simple geometric model to examine how the animal body moved to overcome barriers on a potential energy landscape. We compared successful and failed attempts to reveal what factors among body deformation and body and appendage behaviors contributed to successful self-righting (Rubin et al., 2018).

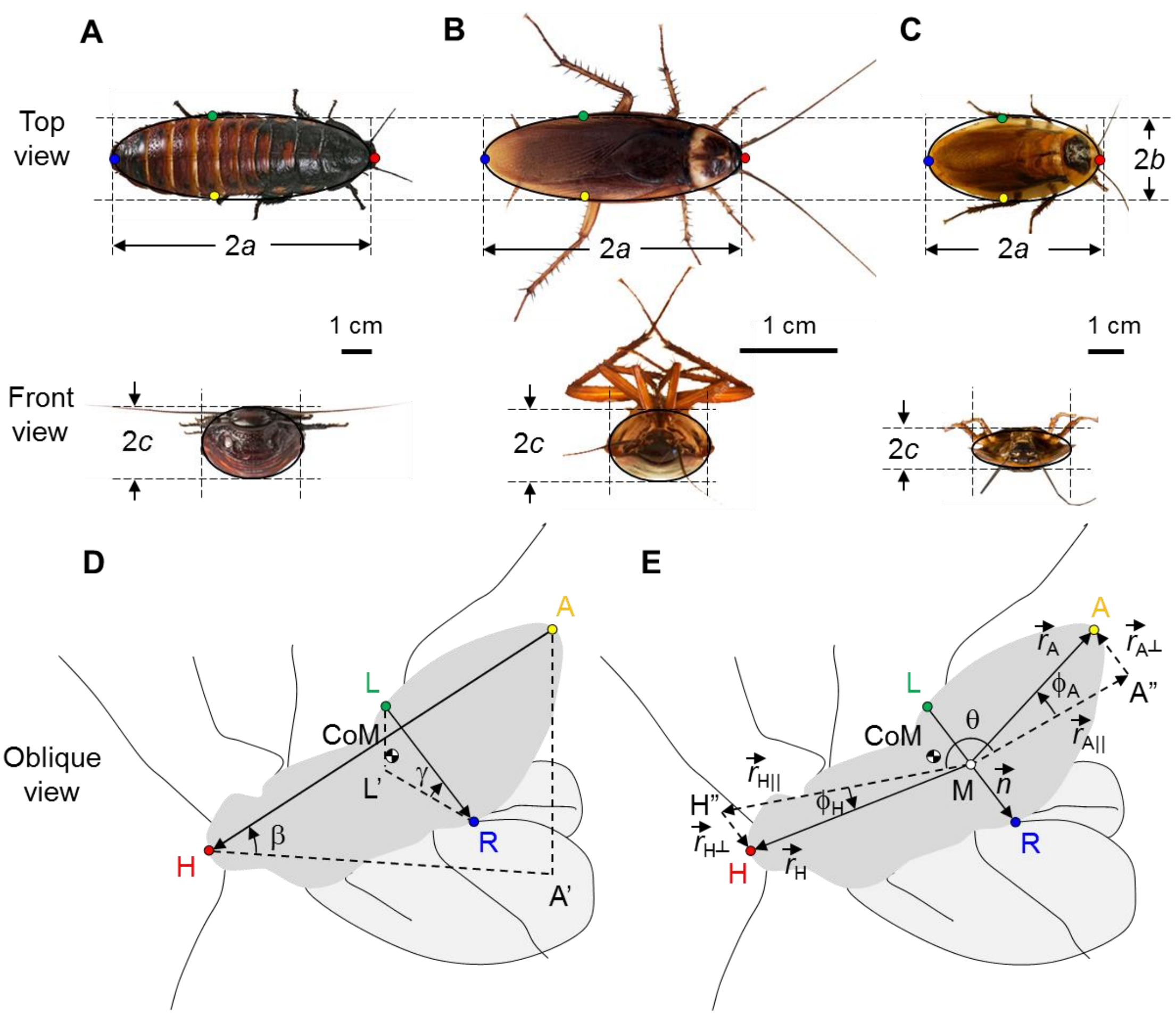

**Fig. 1. Interspecies comparison of body and appendage morphology and schematic defining**

**digitized markers and variables.** (A) Madagascar hissing cockroach. (B) American cockroach. (C) Discoid cockroach. The animals are scaled to the same width in top and front views to illustrate differences in body elongation and flatness (Table 1). The body shape of each species is well approximated by an ellipsoid, with length, width, and thickness of $2a$, $2b$, and $2c$, respectively. Four colored points in the top views are the four digitized markers. (D, E) Schematics of a self-righting animal, showing the four digitized markers, head (H), abdomen (A), left (L), right (R), and definitions of body pitch $\beta$, body roll $\gamma$, body flexion $\theta$, head twisting $\phi_H$, and abdomen twisting $\phi_A$. A' and L' are downward projections of A and L to the same height levels of H and R, respectively. M is a point midway between L and R. $\vec{n}$ is the plane normal of the estimated sagittal plane. H'' and A'' are projections of H and A into the sagittal plane. In the example shown (discoid cockroach using wings), body is flexing, head is twisting to the right, and abdomen is twisting to the left.

**Table 1. Sample size and morphological measurements** (mean ± 1 s.d.).

| Species | Madagascar | American | Discoid |
|---|---|---|---|
| Number of individuals | 6 | 7 | 7 |
| Number of trials | 55 | 59 | 61 |
| Number of successful trials within 30 seconds | 54 | 56 | 58 |
| Number of failed trials within 30 seconds | 1 | 3 | 3 |
| Number of successful trials on first attempt | 41 | 40 | 29 |
| Number of successful trials needing more than one attempt | 13 | 16 | 29 |
| Number of attempts | 78 | 95 | 205 |
| Number of successful attempts | 54 | 56 | 58 |
| Number of failed attempts | 24 | 39 | 147 |
| Mass (g) | 7.44 ± 1.17 | 0.66 ± 0.05 | 2.14 ± 0.15 |
| Body length 2a (cm) | 6.03 ± 0.42 | 3.34 ± 0.14 | 4.98 ± 0.17 |
| Body width 2b (cm) | 2.24 ± 0.10 | 1.19 ± 0.07 | 2.38 ± 0.11 |
| Body thickness 2c (cm) | 1.32 ± 0.10 | 0.70 ± 0.01 | 0.96 ± 0.02 |
| Front leg length (cm) | 2.08 ± 0.08 | 1.62 ± 0.03 | 1.91 ± 0.10 |
| Mid leg length (cm) | 2.93 ± 0.03 | 2.20 ± 0.09 | 2.67 ± 0.06 |
| Hind leg length (cm) | 3.65 ± 0.10 | 3.12 ± 0.03 | 3.60 ± 0.00 |
| Body elongation (body length / body width) | 2.69 ± 0.22 | 2.81 ± 0.20 | 2.09 ± 0.12 |

## MATERIALS AND METHODS

### Animals

We used six Madagascar hissing cockroach, seven American cockroaches, and seven discoid cockroaches. We used adult males because females were often gravid and under different load-bearing conditions. Prior to experiments, we kept the cockroaches in individual plastic containers at room temperature (28 °C) on a 12 h: 12 h light: dark cycle and provided water and food (fruit and dog chow) *ad libitum*. See Table 1 for animal body mass and body and leg dimensions.

The Madagascar hissing and American cockroaches are both relatively elongate and have similar body aspect ratios (body length vs. body width vs. body thickness) (Table 1, Fig. 1A, B). By contrast, the discoid cockroach is less elongate (ANOVA, $P < 0.05$) and flatter (ANOVA, $P < 0.05$) (Table 1, Fig. 1C). The American and discoid cockroach have wings, whereas the Madagascar hissing cockroach are wingless.

### Experimental setup and protocol

We used a low friction, level, flat, rigid surface as the righting arena. The surface was covered with low-friction cardstock, with static friction coefficient $\mu = 0.10 \pm 0.01$ (mean ± 1 s.d.) between the ground and dorsal surface of the animal body (measured by the inclined plane method). Sidewalls around the arena prevented animals from escaping. Four 500W work lights above and three fluorescent lights around the righting arena provided lighting for the high-speed cameras. The temperature during experiments was 36.5 °C. Two webcams (Logitech C920) recorded the entire experiments from top and side views at 30 frame/s. Four synchronized high-speed cameras (AOS and Fastec) recorded up to 30 seconds of each trial from four sides of the arena at 250 frames/s and $800 \times 600$ resolution.

For every trial, we held the animal in an upside-down orientation by grasping the edges of its pronotum and gently released it from a small height (< 0.5 cm) above the center of the area. The small drop was to ensure that the animal did not begin leg searching, a common strategy used for self-righting, before it was set to be upside down on the ground. From high-speed videos, we verified that kinetic energy from the small drop dissipated so that the animal was stationary before it initiated the self-righting response. If the animal did not right within 30 seconds, it was picked up and placed back into its container for rest. We tested all individuals of all three species by alternating individuals and species to ensure sufficient time (> 10 minutes) for each individual to rest between trials to minimize the effect of fatigue (Camhi, 1977).

**Sample size**

Excluding trials in which the animals touched the sidewalls when attempting to self-right, we collected a total of 176 trials from a total of 20 individuals from the three species of cockroaches, with approximately 9 trials from each individual. Because the animal often needed more than one attempt to self-right, from the 176 trials, we identified a total of 378 attempts (see definition below), including 168 successful attempts and 210 failed attempts. See Table 1 for details of sample size.

**Definition of attempts**

Because the animal was allowed up to 30 seconds during each trial, much longer than the time of a typical self-righting attempt (Fig. S1A), the animal may make more than one attempt in a trial. Thus, for each trial, we observed the videos to record how many attempts the animal made, whether each attempt was successful or not, and measured the duration of each attempt.

We defined an attempt as the entire process during which the animal moved its body and appendages to eventually generate a pitching and/or rolling motion, because change in body yaw did not contribute to self-righting. We separated two consecutive attempts by when the animal returned to an upside-down orientation in between the two pitching and/or rolling motions. By this definition, each failed attempt not only included the duration of the body pitching and/or rolling motion, but also the duration prior to it during which the body and appendages moved to generate the attempt. We note that attempts by this definition may and often do include multiple movement cycles of wing opening/closing or leg pushing or flailing, which often occur at higher frequencies than body pitching and/or rolling motion. We did not use wing or leg motion to define attempts because they do not necessarily generate body pitching or rolling, which are defining features towards self-righting.

We then separated attempts into successful and failed ones depending on whether it resulted in self-righting. Each trial can have up to one successful attempt preceded by zero to several failed attempts.

**Performance analysis**

For each trial, we recorded whether the animal succeeded in self-righting within 30 seconds. We also recorded whether the animal succeeded in self-righting on the first attempt of each trial. For each successful trial, we recorded the total number of attempts it took the animal to self-right. We measured total self-righting time, defined as the duration from the instant the animal's dorsal surface touched the surface in an upside-down orientation to the instant when all its six legs touched the ground after the body became upright. We also measured successful attempt time, defined as the duration of the final successful attempt of each successful trial. We calculated the probabilities of self-righting within 30 seconds and on the first attempt, as the ratio of their occurrences to the total number of trials for each species.

**Strategy transition analysis**

To quantify the transitions between strategies during self-righting, we created a locomotor ethogram analysis using each trial (Blaesing, 2004; Li et al., 2015). For each species, we first recorded the sequence of locomotor strategies and the outcome (either successful self-righting or failure). We then calculated the animal's probabilities of entering various self-righting strategies, transitioning between them, and attaining a final outcome. The probability of each transition between nodes was defined as the ratio of the number of occurrences of that transition to the total number of trials of each species. To quantify the often-repeated failed attempts before the final successful attempt, we also counted the number of times the animal continued to use the same strategy consecutively for each trial, and we averaged this number across all the trials of each species to obtain the probability of self-transitions.

**Body rotation and deformation analysis**

To quantify body rotation and deformation during self-righting for each attempt, we digitized four markers on the animal's body (Fig. 1D, E) at the start and end of the attempt and when the body was highest. The instance when the body CoM was highest was determined from high-speed videos by observing when the body stopped pitching and/or rolling upward and began pitching and/or falling downward.

The four markers included: a head marker at the tip of the head (H), an abdomen marker at the tip of the abdomen (A), a left marker on the left side of the abdomen (L), and a right marker on the right side of the abdomen (R). Both the left and right markers were located at about 60% body length from the head, close to the fore-aft position of the center of mass (Kram et al., 1997). Each marker was digitized in at least two high-speed videos from different views using DLTdv5 (Hedrick, 2008), which were used to reconstruct 3-D positions using DLTcal5 (Hedrick, 2008) and a custom 27-point calibration object. The position midway (M) between the left and right markers was calculated.

We approximated the CoM position using the average position of all four markers. Using positions of the tips of the head (H) and abdomen (A), we calculated body pitch and body yaw relative to the ground. Using positions of the left (L) and right (R) points on the sides of the abdomen, we calculated body roll relative to the ground. In addition, we calculated body flexion $\theta$ as the angle within the sagittal plane formed between the in-plane components ($\vec{r}_{H\parallel}$ and $\vec{r}_{A\parallel}$) of two vectors $\vec{r}_H$ and $\vec{r}_A$, which started from the midway point (M) and pointed to the head (H) and abdomen (A) markers, respectively. $\vec{r}_{H\perp}$ and $\vec{r}_{A\perp}$ are components of $\vec{r}_H$ and $\vec{r}_A$ perpendicular to the sagittal plane. A negative body flexion meant body hyperextension. Further, we calculated head and abdomen twisting, $\phi_H$ and $\phi_A$, as the angles between the sagittal plane and the vectors $\vec{r}_H$ and $\vec{r}_A$, respectively. Sagittal plane was approximated by a plane whose normal vector $\vec{n}$ was the vector from the left (L) to the right (R) marker. See Fig. 1D, E for details. Equations are summarized below.

CoM position:

$$x_{CoM} = 1/4(x_H + x_A + x_L + x_R)$$

$$y_{CoM} = 1/4(y_H + y_A + y_L + y_R)$$

$$z_{CoM} = 1/4(z_H + z_A + z_L + z_R)$$

body orientation:

$$\text{pitch} = \tan^{-1}[(z_A - z_H)/\text{sqrt}((x_A - x_H)^2 + (y_A - y_H)^2]$$

$$\text{roll} = \tan^{-1}[(z_L - z_R)/\text{sqrt}((x_L - x_R)^2 + (y_L - y_R)^2]$$

$$\text{yaw} = \tan^{-1}[(y_A - y_H)/(x_A - x_H)]$$

body flexion:

$$\theta = \cos^{-1}[(\vec{r}_{H\parallel} \bullet \vec{r}_{A\parallel})/(|\vec{r}_{H\parallel}|\ |\vec{r}_{A\parallel}|)]$$

head twisting:

$$\phi_H = \tan^{-1}(|\vec{r}_{H\perp}|/|\vec{r}_{H\parallel}|)$$

abdomen twisting:

$$\phi_A = \tan^{-1}(|\vec{r}_{A\perp}|/|\vec{r}_{A\parallel}|)$$

where:

$$\vec{n} = (x_L, y_L, z_L) - (x_R, y_R, z_R)$$

$$\vec{r}_{H\perp} = \vec{r}_H \bullet \vec{n}$$

$$\vec{r}_{H\parallel} = \vec{r}_H - \vec{r}_H \bullet \vec{n}$$

$$\vec{r}_{A\perp} = \vec{r}_A \bullet \vec{n}$$

$$\vec{r}_{A\parallel} = \vec{r}_A - \vec{r}_A \bullet \vec{n}$$

To study how the animal moved in an attempt to self-right, we calculated the changes in each of these variables from the start of each attempt to when the body CoM was highest. When doing this, we used absolute values of pitch, roll, and yaw considering symmetry of the animal's ellipsoid-shaped body. We also set head and abdomen twisting at the start of the attempt to be always non-negative, considering lateral symmetry of the animal.

**Body and appendage behavior analysis**

To further identify what contributed to successful self-righting, for each attempt, we recorded the following events (or lack thereof) to quantify the animal's body and appendage behaviors:

(1) Dynamic: whether the animal's body rotation was dynamic, being able to gain and use pitch and/or roll rotational kinetic energy to overcome potential energy barriers. Dynamic behavior was determined by observing whether the animal's body was still moving upward when its appendage used for self-righting (wings or legs) had stopped pushing against the ground. A wing stopped pushing against the ground when its distal section lifted off the ground as the body pitched and/or rolled. A leg stopped pushing against the ground when its distal segments, which engaged the surface for self-righting, slipped, reducing vertical force production (Full et al., 1995). When either of these occurred, the body could only continue to move upward if it still had rotational kinetic energy.

(2) Body lift-off: whether the body lifted off from the surface.

(3) Body hold: whether the body was held in the air after pitching up so that the abdomen remained raised, when using wings to self-right.

(4) Body sliding: whether the animal's body slid on the ground as it pitched/rolled toward self-righting.

(5) Leg assist: whether legs assisted by pushing against the surface to generate body pitching and/or rolling towards self-righting, when using wings to self-right.

(6) Leg slip: whether the leg engaging the surface to self-right (both as the primary and assisting mechanisms) slipped on the surface.

(7) Accelerate: whether the assisting leg accelerated body pitching and/or rolling motion towards self-righting.

(8) Overshoot: whether there was any overshooting in body pitching and/or rolling motion beyond the upright orientation that must be corrected by legs.

We calculated the probabilities of these body and appendage behaviors as the ratios of the occurrences of each to the total number of attempts for each strategy, separated by whether the attempt was successful or not.

All data analyses were performed using Microsoft Excel and MATLAB.

**Statistics**

Before pooling trials, for each species, we performed a mixed-design ANOVA (for continuous variables) or a chi-square test (for nominal variables), both with trial number as a fixed factor and individual as a random factor to account for individual variability. We found no effect of trial for any measurements relevant to a trial ($P > 0.05$, ANOVA or $P > 0.05$, chi-square test), including number of attempts to self-right, self-righting probabilities, righting times, and transition probabilities. Thus, we pooled all trials from each individual to calculate their means and confidence intervals (for nominal variables) or standard deviations (for continuous variables).

Before pooling attempts, for each species, we performed a mixed-design ANOVA (for continuous variables) or a chi-square test (for nominal variables), both with attempt number as a fixed factor and individual as a random factor to account for individual variability. We found no effect of attempt for most (72 out of 84) measurements relevant to an attempt ($P > 0.05$, ANOVA or $P > 0.05$, chi-square test), including attempt time, changes in body pitch, roll, yaw, CoM height, body flexion, head and abdomen twisting, and body and leg behavior probabilities. Thus, we pooled all attempts for each of the self-righting strategies, separated by whether the attempt was successful or not, to calculate their means and confidence intervals (for nominal variables) or standard deviations (for continuous variables).

To test whether measurements relevant to an attempt differed between successful and failed attempts, for each species using each strategy, we used a mixed-design ANOVA (for continuous variables) or a chi-square test (for nominal variables), with the successful/failure record as a fixed factor and individual as a random factor to account for individual variability.

To test whether measurements relevant to the strategy used (winged or legged) differed between winged and legged attempts, for each species separated by whether the attempt was successful or not, we used a mixed-design ANOVA or a chi-square test (for nominal variables), with the strategy used as a fixed factor and individual as a random factor to account for individual variability.

To test whether measurements relevant to successful trials differed between species, we used a mixed-design ANOVA (for continuous variables) or a chi-square test (for probabilities), with species as a fixed factor and individual as a nested, random factor to account for individual variability.

Wherever possible, we used Tukey's honestly significant difference test (HSD) to perform post-hoc analysis. All the statistical tests followed (McDonald, 2009) and were performed using JMP.

**Potential energy landscape model using simple body geometry**

To visualize how the animal rotated during self-righting attempts and how this differed between strategies and species, we used a simple geometric model to calculate the potential energy landscape of the body (Fig. 2). Because the animal rarely lifted off the ground during self-righting for all three species (7 out of 378 attempts, Fig. S2B), as a first-order approximation, we considered the animal body as an ellipsoid with its lowest point in contact with a horizontal, flat surface (Fig. 2A-C). Ellipsoid length $2a$, width $2b$, and thickness $2c$ were body length, width, and thickness from morphological measurements (Table 1). We approximated the CoM position with the ellipsoid's geometric center (Kram et al., 1997).

The simple geometric model allowed us to visualize the state of an ellipsoidal body on a potential energy landscape (Fig. 2D). For an elongate ellipsoid body, self-righting by pitching overcomes the highest potential energy barrier (Fig. 2A), whereas self-righting by rolling overcomes the lowest barrier (Fig. 2B). Self-righting by a diagonal body rotation (Frantsevich, 2004), with simultaneous pitching and rolling, overcomes an intermediate barrier (e.g., Fig. 2C, an ideal diagonal rotation about a fixed axis in the horizontal plane between pitch and roll axes). Body yawing did not affect CoM position or barrier height (because we used the yaw-pitch-roll convention of Euler angles).

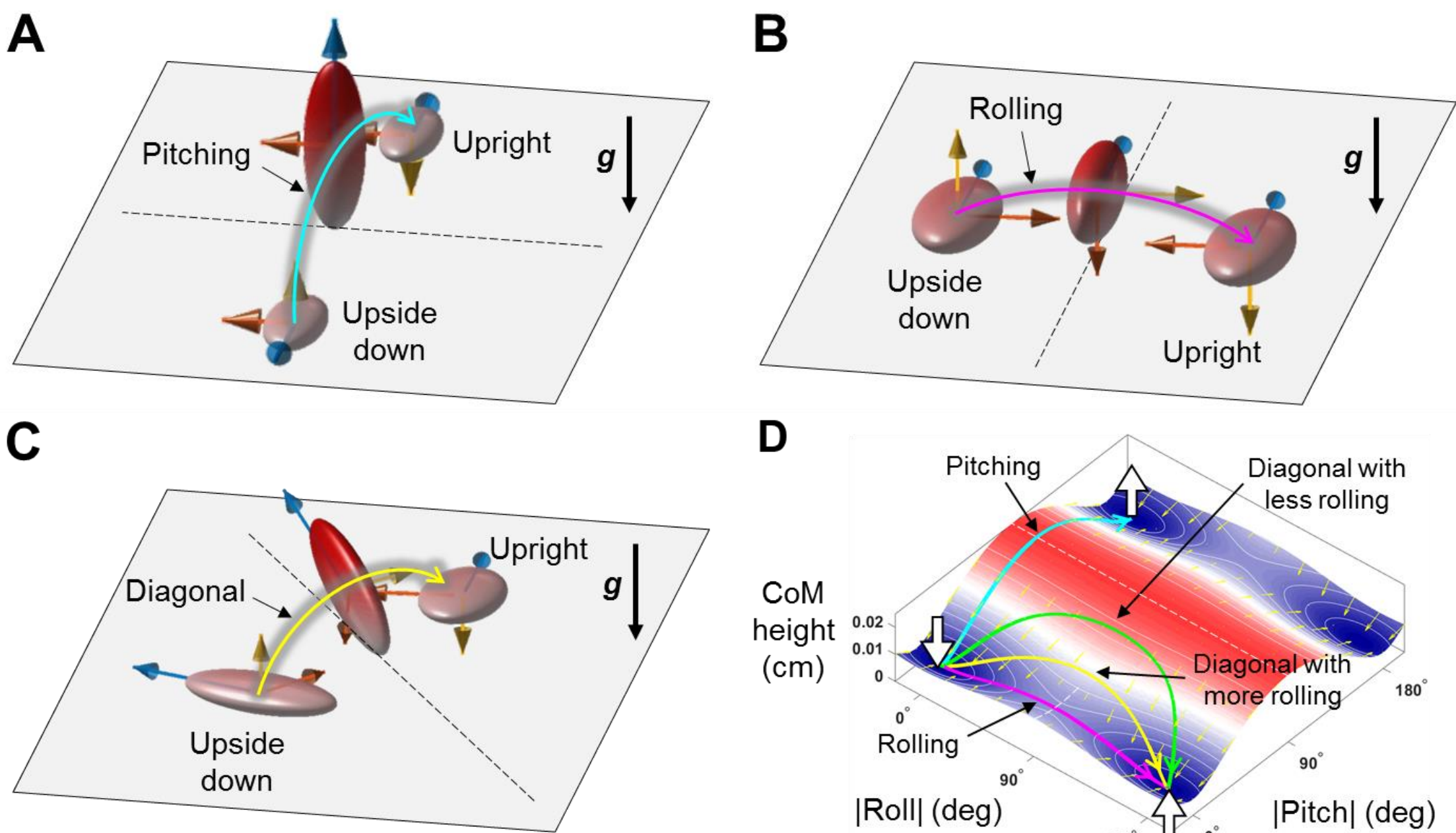

**Fig. 2. Potential energy landscape from the simple geometric model.** (A-C) An ellipsoid approximating the animal body in contact with the ground, either pitching (A), rolling (B), or rotating diagonally (simultaneous pitching and rolling) (C). Dashed line shows rotation axis.

Diagonal rotation shown is about a fixed axis within the horizontal ground plane for simplicity; actual diagonal rotation of the animal may be about a time-varying axis. Red, blue, and yellow arrows on each ellipsoidal body show its three major axes to illustrate body rotation. Vector $g$ shows the direction of gravity. (D) Potential energy landscape, shown as CoM height as a function of body pitch and body roll (using Euler angles with yaw-pitch-roll convention), calculated from the geometric model. We use absolute values of body pitch and roll considering symmetry of the ellipsoid. Downward and upward arrows indicate an upside-down and upright body orientation, respectively. Cyan, green, yellow, and magenta curves with arrows are representative trajectories for pure pitching, two different diagonal rotations, and pure rolling, all about a fixed axis in the horizontal plane, to illustrate the fact that more body rolling decreases the potential energy barrier. White curves on the landscape are iso-height contours. Small yellow arrows on the landscape are gradients. Model results shown are using the discoid cockroach's body dimensions as an example.

## RESULTS

### Self-righting attempts

For all three species, self-righting on a flat, rigid, low friction surface was a challenging task and often required more than one attempt to succeed (Table 1, Fig. 3A; 24%, 29%, and 48% of all trials had multiple attempts for the Madagascar hissing, American, and discoid cockroaches, respectively). Repeated attempts were consistent with previous observations in the discoid cockroach (Full et al., 1995). The Madagascar hissing, American, and discoid cockroaches needed an average of 1.3, 1.8, and 3.2 attempts to self-right. The difference was significant only between the Madagascar and discoid cockroaches ($P < 0.05$, ANOVA, Tukey HSD).

For all three species, we found no dependence on trial number or only a few cases of dependence on attempt number (see **Statistics**). The lack of dependence on trial number showed that there was only a minor effect of history dependence on self-righting and that the animal's motion and use of strategies was stochastic and unpredictable (Full et al., 1995) over consecutive attempts.

### Self-righting probability

All three species self-righted with high probability when given time (30 seconds in our experiments; Fig. 3B, white bar; averaging 97% for all three species) and self-righted on the first attempt in over half of all trials (Fig. 3B, gray; averaging 63% for all species) with no significant difference across species ($P > 0.05$, chi-square test).

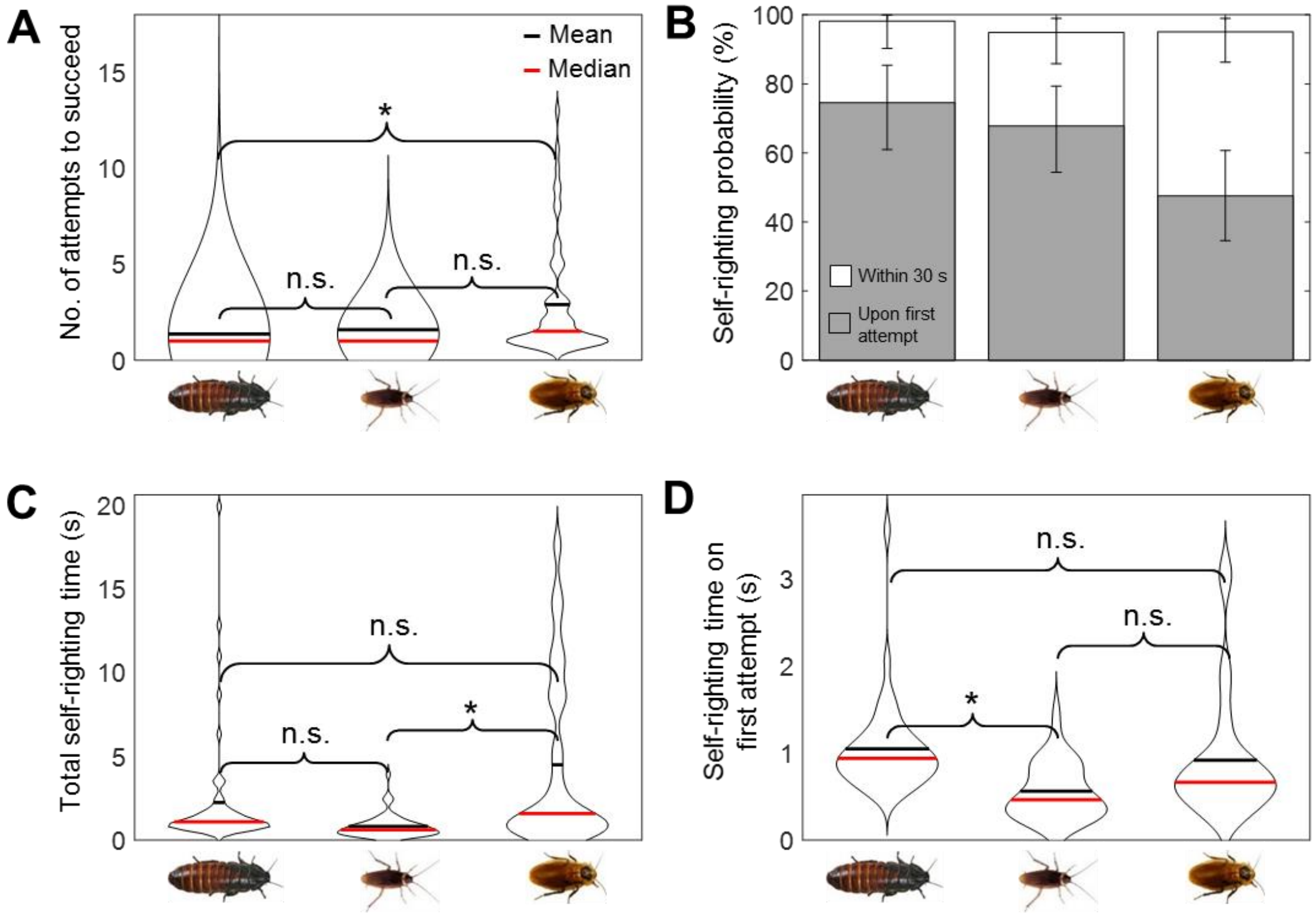

**Fig. 3. Self-righting performance.** (A) Number of attempts to achieve self-righting. *n* = 55, 59, 61 for (B) Self-righting probability within 30 seconds (white) and on the first attempt (gray). Error bars represent 95% confidence intervals. (C) Total time to achieve self-righting. (D) Time to achieve self-righting on the first attempt. From left to right are Madagascar hissing, American, and discoid cockroaches. In A, C, and D, data are shown using violin plots. Black and red lines show the mean and median, respectfully. Width of graph shows the frequency of the data along the *y*-axis. Brackets show whether there is a significant difference (asterisk, $P < 0.05$, ANOVA) or not (n.s.). See Table 1 for sample size.

### Self-righting time

All three species were capable of self-righting rapidly. The fastest self-righting took only 0.14 s for the American cockroach, 0.31 s for the discoid cockroach, and 0.46 s for the Madagascar hissing cockroach. The median total time to achieve self-righting including failed attempts was 1.1 s, 0.6 s, and 1.6 s for the Madagascar hissing, American, and discoid cockroaches, respectively (Fig. 3C). The maximal time was 19.9 s, 3.9 s, and 17.7 s for the Madagascar hissing, American and discoid cockroaches, respectively. The difference was only significant between the American and discoid cockroaches ($P < 0.05$, ANOVA, Tukey HSD). The mean self-righting time on the first attempt (Fig. 3D) was 1.0 s for the Madagascar hissing cockroach, longer than the American cockroach's 0.6 s ($P < 0.05$, ANOVA, Tukey HSD), although neither differed from the discoid cockroach's 0.9 s.

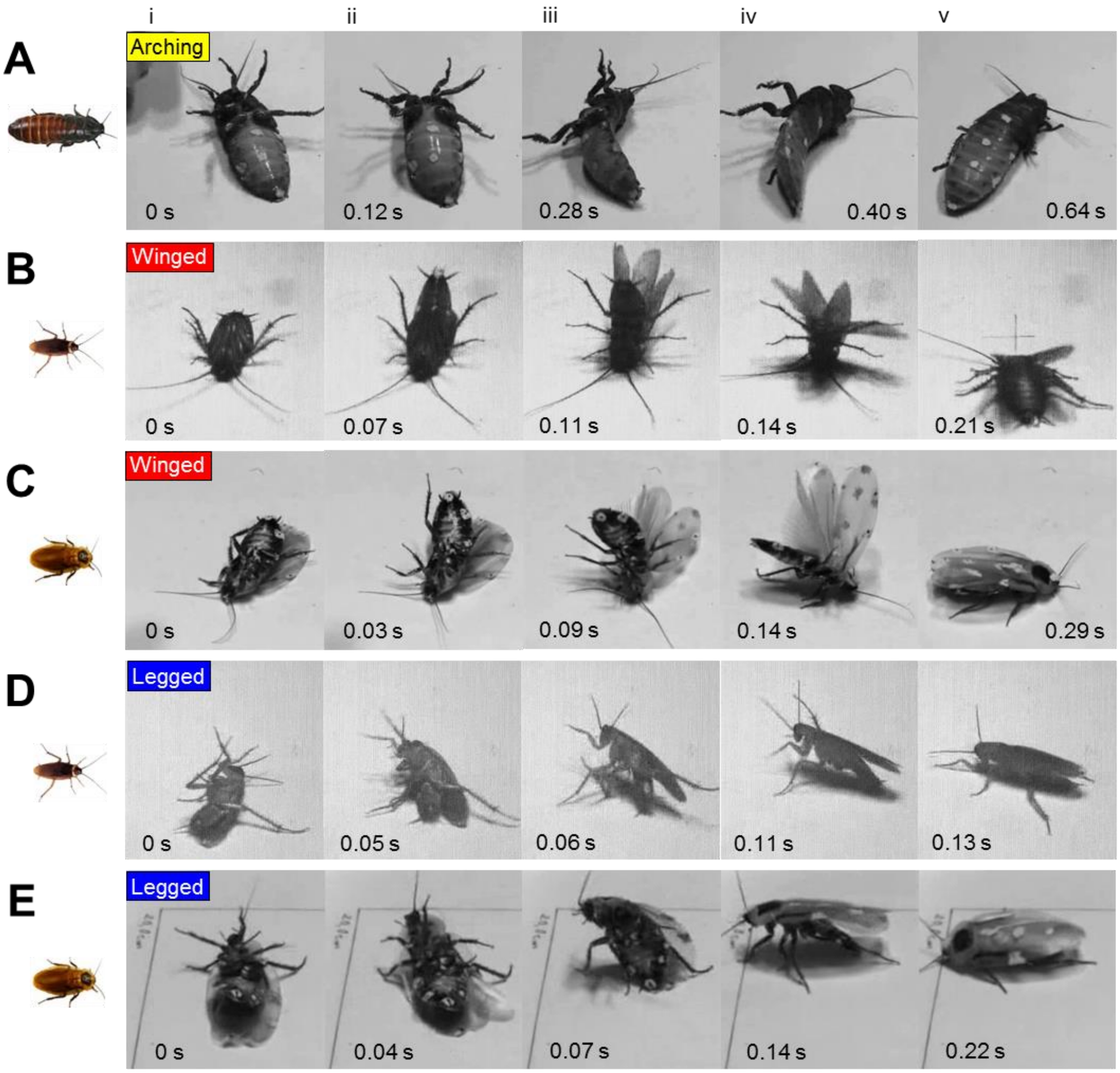

**Fig. 4. Representative snapshots of self-righting strategies.** (A) Madagascar hissing cockroach using body arching. (B) American cockroach using wings. (C) Discoid cockroach using wings. (D) American cockroach using legs. (E) Discoid cockroach using legs. i-v are five snapshots forward in time.

### Self-righting strategies

**Body arching.** The Madagascar hissing cockroach's self-righting relied primarily on changing body shape assisted by legs (Fig. 4A, Movie 1). When lying upside down (Fig. 4A, i), the animal hyperextended its body into an arch to raise the CoM (Fig. 4A, ii) (Camhi, 1977), similar to some beetles (Frantsevich, 2004). The narrow static stability region between the head and tip of the abdomen in contact with the ground and lateral perturbations from flailing legs induced the body to roll (Fig. 4A, iii). As the body fell onto one side, rolling stopped due to resistance from the legs and the metastable body shape in the transverse plane (Camhi, 1977), resembling that of medium-height turtle shells (Domokos and Várkonyi, 2008). Then, the legs on the lowered side kept pushing, resulting in skidding and yawing on the surface, while the body continued hyperextending (Fig. 4A, iv). When a body arching attempt failed, the animal sometimes quickly flexed its body straight (occurring at a 25% probability per attempt) to reverse the direction of body rolling using rotational kinetic energy gained due to falling of the CoM to start another body arching attempt. When one of the pushing legs eventually managed to wedge under the body, its thrust rolled the body further over protruding legs to overcome their secondary potential energy barriers to achieve self-righting (Fig. 4A, v).

**Wing use.** Both the American and discoid cockroaches can self-right primarily using wings (Fig. 4B, C, Movie 2). When lying upside down (Fig. 4B, C, i), the animal separated its wings laterally and pronated them so that their outer edges pushed against the surface while the head remained in contact as a pivot, which pitched the abdomen upward (Fig. 4B, C, ii) and often resulted in additional body rolling. When a winged attempt failed, the animal closed its wings to pitch back downward and sometimes started the same process again in another attempt (occurring at a 3% probability per attempt for the American cockroach and an average of 1.1 times per attempt for the discoid cockroaches). When a winged attempt succeeded, the animal fell with additional body pitching and/or rolling to become upright (Fig. 4B, C, iii, iv, v). Legs flailed in this process, resulting in small lateral perturbations. Flailing legs frequently hit and pushed against the ground (91% of attempts) providing impulses to change body rotation.

**Leg use.** The American and discoid cockroaches can also self-right primarily using legs (Fig. 4D, E, Movie 3). When lying upside down, these insects always continuously kicked their legs outward in an attempt to push against the ground (Reingold and Camhi, 1977; Zill, 1986). Frequent legs slipping (55% of attempts) due to the low friction of the surface resulted in continuous body sliding (41% of attempts). In failed attempts, body rolling and pitching induced by kicking legs were too small to achieve self-righting, and the animal started the same process in another attempt (occurring at a 51% and 21% probability per attempt for the American and discoid cockroaches, respectively). When a legged attempt succeeded (Fig. 4D, E, i), two legs engaged the surface simultaneously (Fig. 4D, E, ii), typically a hind leg and a contralateral middle leg (76% and 93% of attempts for the American and all attempts for the discoid cockroaches, respectively). The two legs pushed to thrust the body forward, pitched it head up, and rolled it such that the abdomen cleared the surface to self-right (Fig. 4D, E, iii, iv, v).

**Probability of dynamic self-righting**

For both the American and discoid cockroaches using both wings and legs, self-righting attempts were often dynamic (Fig. 5; American: 67% of winged attempts, 55% of legged attempts, 56% of all attempts; discoid: 37% of winged attempts, 80% of legged attempts, 51% of all attempts), being able to gain and use pitch and/or roll rotational kinetic energy in an attempt to overcome potential energy barriers. By contrast, the Madagascar hissing cockroach's self-righting using body arching was never dynamic (0%, Fig. 5).

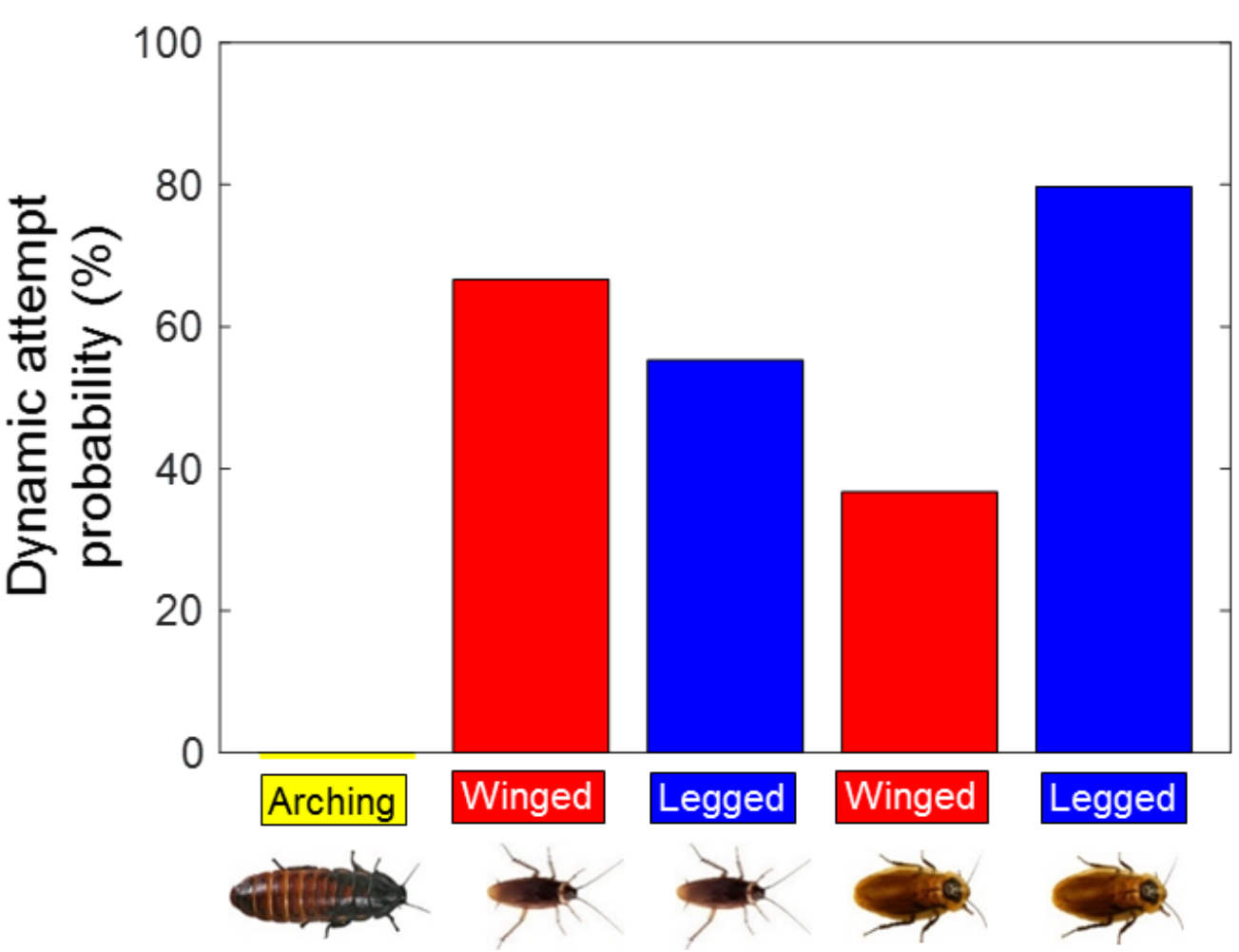

**Fig. 5. Probability of self-righting dynamically.** A dynamic attempt is one in which the animal is able to gain and use pitch and/or roll rotational kinetic energy in an attempt to overcome potential energy barriers, whether the attempt is successful or not. See Table 1 for sample size.

**Self-righting transitions**

All three species attempted more than one strategy and often transitioned between them to self-right, even though not all of them led to successful righting on the flat, rigid, low-friction surface (Fig. 6). Both the Madagascar hissing and American cockroaches' self-righting was more stereotypical and primarily used one successful strategy, in contrast to the discoid cockroach that used two successful strategies nearly equally.

The Madagascar hissing cockroach (Fig. 6A) most frequently used body arching to self-right (85% of all trials). When not successful, this cockroach always continued to use body arching, leading to a high probability of self-righting (98%). It occasionally used body twisting (13%) (Camhi, 1977) which never succeeded and after which it always transitioned to body arching (13%).

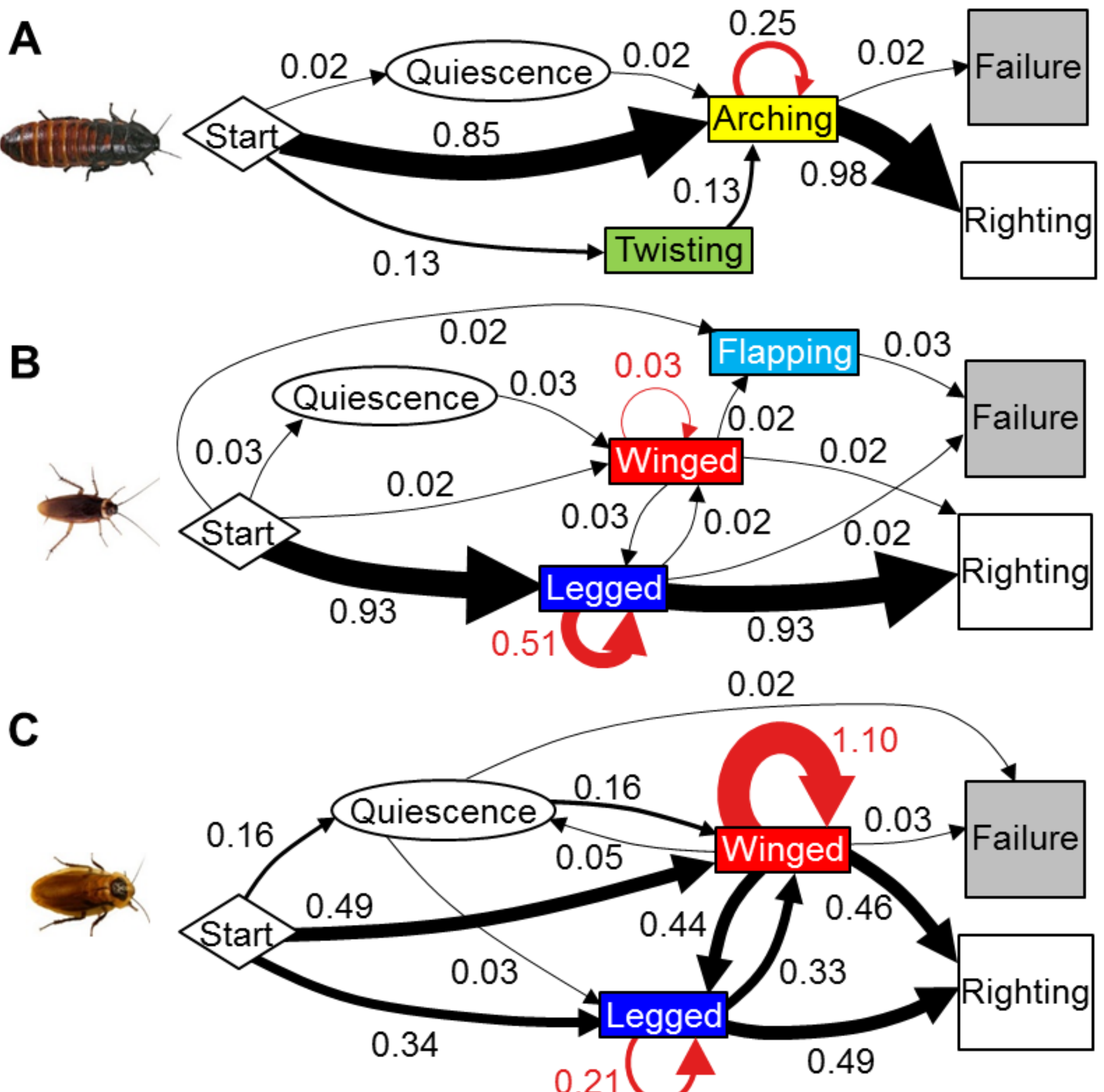

**Fig. 6. Self-righting locomotor transition ethograms.** (A) Madagascar hissing cockroach. (B) American cockroach. (C) Discoid cockroach. Arrow widths are proportional to transition probabilities between nodes, with probability values shown by numbers. Transition probabilities are defined as the ratio of the number of occurrences of each transition to the total number of trials for each species. Red arrows and numbers are for self-transition and represent the average number of times of continuing the same strategy during each trial. A self-transition probability greater than one means that on average it occurred more than once for each trial. The sum of transition probabilities out of each node equals that into the node, except for start with a total probability of 1 going out, and righting and failure with a total probability of 1 into both together. See Table 1 for sample size.

The American cockroach (Fig. 6B) most frequently used legs (93%) and occasionally used wings (2%), despite being capable of self-righting using both strategies. When not successful, it often continued to use the same legged or winged strategy, but also occasionally transitioned between them. It also infrequently used flapping (2%) which never succeeded.

By contrast, the discoid cockroach (Fig. 6C) initially used either wings (49%) or legs (34%) to self-right. When unsuccessful, it continued to use the same legged or winged strategy, but also frequently transitioned between them, resulting in high probabilities of self-righting (46% or 49% eventually using wings or legs to self-right, respectively).

All three species occasionally entered a temporary quiescence mode (Camhi, 1977) without apparent body or appendage movement (2%, 3%, and 16% for Madagascar hissing, American, and discoid cockroaches, respectively). The Madagascar cockroach occasionally used body twisting (13%) and the American cockroach showed wing flapping (2%) in an attempt to self-right, but these two strategies never succeeded.

**Body state on potential energy landscape**

For all three species, because the body rarely lifted off the ground for all three species (7 out of 378 attempts, Fig. S2B), the measured state of the animal (body pitch, body roll, and CoM height) lied on the surface of the potential energy landscape using the simple geometric model (Fig. 7). Being on the surface of the energy landscape allowed us to examine how the animal's body moved through three stages (start, highest CoM height, and end) of an attempt to overcome potential energy barriers (or lack thereof).

For the Madagascar hissing cockroach using body arching and the American cockroach using legs, body rotation was mainly rolling during both successful and failed attempts (Fig. 7A, i, ii; Fig. 7C, i, ii), which overcame the lowest potential energy barrier if successful (Fig. 7A, i). For the American cockroach using wings, body rotation was mainly pitching during both successful and failed attempts (Fig. 7B, i, ii), which overcame the highest potential energy barrier if successful (Fig. 7B, i). For the discoid cockroach using both wings and legs, body rotation involved simultaneous pitching and rolling during successful attempts (Fig. 7D, i; Fig. 7E, i), which overcame intermediate potential energy barriers, and body rotation was mainly pitching during failed attempts (Fig. 7D, ii; Fig. 7E, ii). In failed attempts, the animal was unable to overcome the potential energy barriers (Fig. 7A-E, ii).

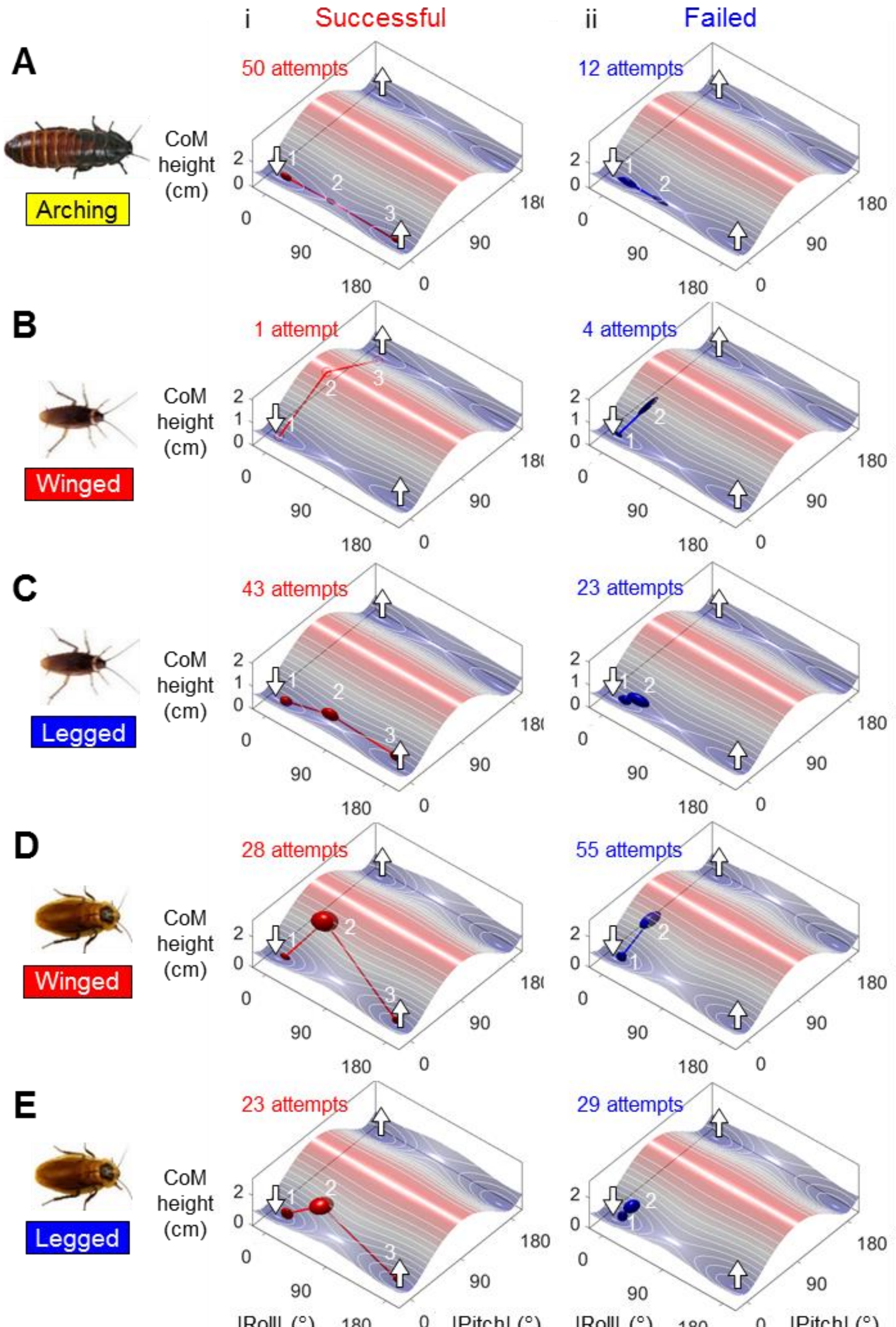

**Fig. 7. State of the body on the potential energy landscape at the start (1), highest CoM position (2), and end (3) of the attempt during successful (i) vs. failed (ii) attempts.** (A) Madagascar hissing cockroach using body arching. (B) American cockroach using wings. (C)

American cockroach using legs. (D) Discoid cockroach using wings. (E) Discoid cockroach using legs. Landscape is defined in Fig. 2D. On each landscape, the ellipsoids show means (center of ellipsoid) ± 1 s.d. (principal semi-axis lengths of ellipsoid) of body pitch, body roll, and CoM height at each stage of the attempt. For failed attempts (ii), the end state (3) is not shown because it nearly overlaps with the start state (1). Sample size of each case is shown. Note that the sample size here combined for each species is slightly smaller than its total number of attempts, because in some attempts the animal markers are out of the field of view and cannot be digitized.

In addition, both the Madagascar hissing and American cockroaches had a large number of successful attempts (50 and 43, respectively) using strategies (body arching and legged, respectively) that overcame low potential energy barriers (Fig. 7A, C, i). The American cockroach had only one successful attempt using wings which overcame high potential energy barriers (Fig. 7B, i). By contrast, the discoid cockroach had similar numbers of successful attempts to overcome potential energy barriers using the two strategies, winged (28%) and legged (23%) self-righting (Fig. 7D, E, i).

**Body rotation and center of mass height increase**

**Madagascar hissing cockroach.** Using body arching to self-right, the Madagascar hissing cockroach pitched little towards 90° (Fig. 8A, i) but rolled substantially towards 90° (Fig. 8B, i) as the body attained its highest CoM position. Rolling resulted in a small CoM height increase relative to the highest potential barrier height possible ($a - c$) (Fig. 8C, i). The body rolled more in successful than in failed attempts ($\Delta|\text{roll}| = 69°$ vs. $50°$; $P < 0.05$, ANOVA).

**American cockroach.** Using wings to self-right, the American cockroach pitched substantially towards 90° (Fig. 8A, ii) and rolled little towards 90° (Fig. 8B, ii) as the body attained its highest CoM position. This resulted in a large CoM height increase relative to the highest potential barrier height possible ($a - c$) (Fig. 8C, ii).

Using legs to self-right, the American cockroach pitched little towards 90° (Fig. 8A, iii) and rolled substantially towards 90° (Fig. 8B, iii) as the body attained its highest CoM position. This resulted in a small CoM height increase relative to the highest potential barrier height possible ($a - c$) (Fig. 8C, iii). The body rolled more in successful than in failed attempts ($\Delta|\text{roll}| = 61°$ vs. $20°$; $P < 0.05$, ANOVA).

For successful attempts, the American cockroach pitched more ($\Delta|\text{pitch}| = 78°$ vs. $5°$; $P < 0.05$, ANOVA), rolled less ($\Delta|\text{roll}| = 6°$ vs. $61°$; $P < 0.05$, ANOVA), and its CoM height increased

more ($\Delta z_{CoM}$ = 1.8 cm vs. 0.7 cm; $P < 0.05$, ANOVA) when using wings than when using legs (Fig. 8A-C, ii vs. iii).

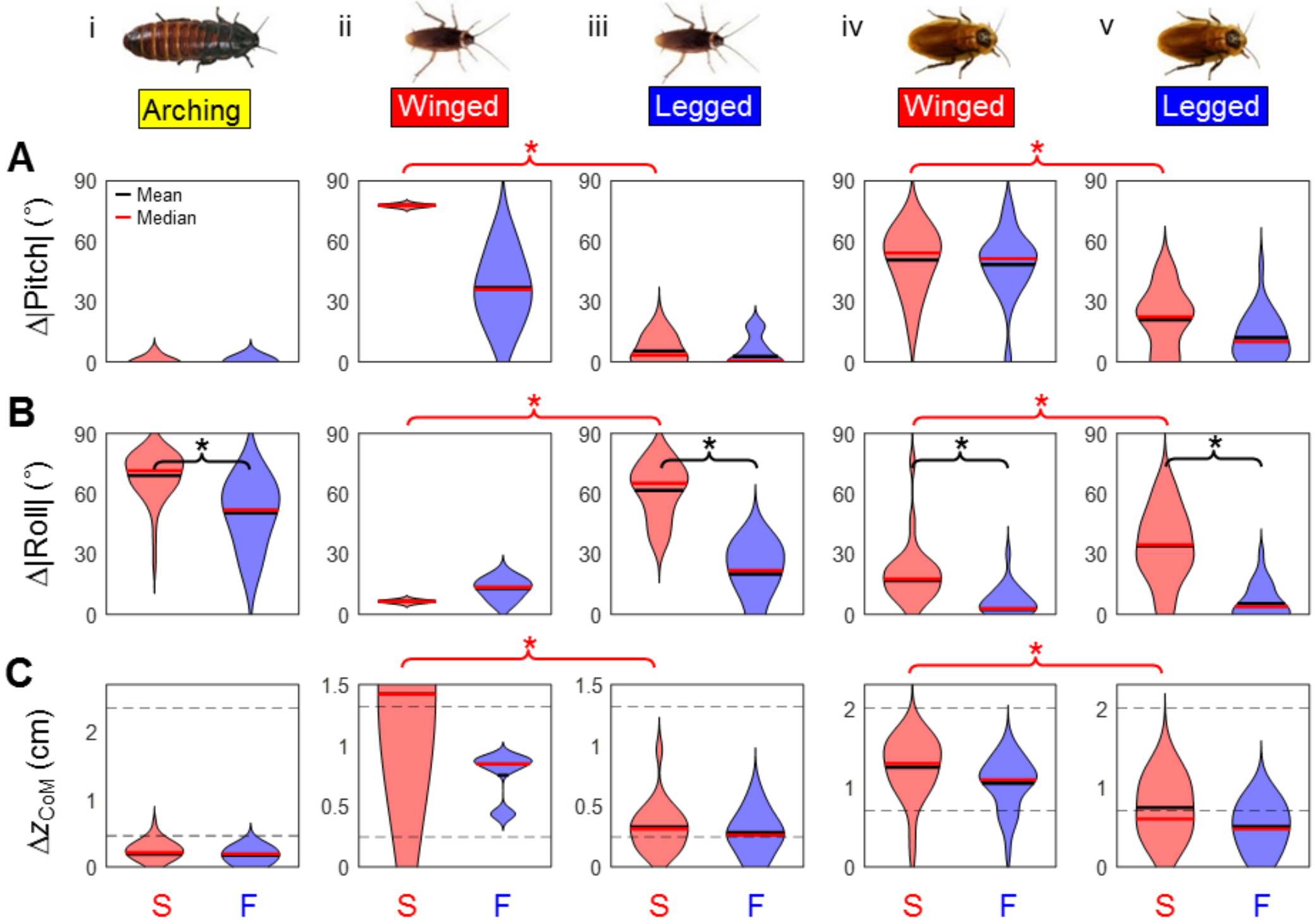

**Fig. 8. Body pitch increase (A), body roll increase (B), and CoM height increase (C) when the body was highest for successful (S, red) vs. failed (F, blue) self-righting attempts.** (i) Madagascar hissing cockroach using body arching. (ii) American cockroach using wings. (iii) American cockroach using legs. (iv) Discoid cockroach using wings. (v) Discoid cockroach using legs. We used absolute values of body pitch and roll considering symmetry of the ellipsoid representing the body. Data are shown using violin plots. Black and red lines show the mean and median. Width of graph shows the frequency of the data along the *y*-axis. Black asterisks and brackets indicate a significant difference between successful and failed attempts ($P < 0.05$, ANOVA). Red asterisks and brackets indicate a significant difference between winged and legged attempts for the same species ($P < 0.05$, ANOVA). In (C), two horizontal dashed lines show the lowest and highest barriers from the ellipsoid model, $b - c$ and $a - c$, for pure rolling and pure pitching, respectively. For successful attempts, CoM height increase is the measured barrier height. For failed attempts, barrier height is not measured because the animal did not overcome it. See Table 1 for sample size.

**Discoid cockroach.** Using wings to self-right, the discoid cockroach pitched substantially towards 90° (Fig. 8A, iv) and rolled less than it pitched (Fig. 8B, iv) towards 90° as the body attained its highest CoM position. This resulted in a large CoM height increase relative to the highest potential barrier height possible ($a - c$) (Fig. 8C, iv). The body rolled more in successful than in failed attempts ($\Delta|\text{roll}|$ = 17° vs. 2°; $P < 0.05$, ANOVA).

Using legs to self-right, the discoid cockroach both pitched (Fig. 8A, v) and rolled (Fig. 8B, v) a little towards 90° as the body attained its highest CoM position. This resulted in a small CoM height increase relative to the highest potential barrier height possible ($a - c$) (Fig. 8C, v). The body rolled more in successful than in failed attempts ($\Delta|\text{roll}|$ = 34° vs. 5°; $P < 0.05$, ANOVA).

For successful attempts, the discoid cockroach pitched more ($\Delta|\text{pitch}|$ = 51° vs. 21°, $P < 0.05$, ANOVA), rolled less ($\Delta|\text{roll}|$ = 17° vs. 34°, $P < 0.05$, ANOVA), and its CoM height increased more ($\Delta z_{\text{CoM}}$ = 1.3 cm vs. 0.8 cm, $P < 0.05$, ANOVA) when using wings than when using legs (Fig. 8A-C, iv vs. v).

**All three species.** For all three species, body rolling was the best predictor of whether an attempt succeeded or failed. Roll increase when the CoM was highest was greater in successful than in failed attempts for all the cases ($P < 0.05$, ANOVA; Fig. 8B, i, iii-v), except for the American cockroach using wings (Fig. 8B, ii) which had a small sample size (1 successful and 4 failed attempts).

Because CoM height increase was the measured potential energy barrier height for successful attempts, both the American and discoid cockroaches overcame higher barriers when using wings than when using legs, and this difference was greater for the American cockroach.

### Other factors contributing to successful self-righting

Besides body rolling, three factors were important in differentiating successful from failed attempts (Fig. S2). First, except for the American cockroach using wings, leg slip was less frequent in successful attempts for all three species (Fig. S2A, i, iii-v; Madagascar arching: successful: 0%, failed: 100%; American legged: successful: 0%, failed: 90%; discoid winged: successful: 45%, failed: 97%; discoid legged: successful: 0%, failed: 100%; $P < 0.05$, chi-square test). Second, for the American cockroach using legs and the discoid cockroach using both wings and legs, legs more frequently hit the ground in successful attempts to accelerate body rotation, after wings or legs generated the initial body pitching and/or rolling (Fig. S2B, iii-v; American legged: successful: 51%, failed: 10%; discoid winged: successful: 62%, failed: 5%; discoid legged: successful: 34%, failed: 0%; $P < 0.05$, chi-square test). Third, for both the American and discoid cockroaches using wings, the body was held in the air with the abdomen pitched upward less frequently in successful attempts (Fig. S2C, ii, iv; American winged: successful: 0%, failed: 80%; discoid winged: successful: 52%, failed: 98%; $P < 0.05$, chi-square test). Body holding was not observed in the legged and arching strategies.

We did not observe significant differences between successful and failed attempts that were consistent across species and strategies for all other measurements (Figs. S1, S3), including attempt time, body yaw change, body flexion change, head and abdomen twisting changes, dynamic probability, body lift-off probability, body sliding probability, leg assist probability, and overshoot probability. We did find significant differences between successful and failed attempts in attempt time for the discoid cockroach using both wings and legs (Fig. S1A, iv, v), in body yaw change for the American cockroach using wings (Fig. S1B, ii), in both head and abdomen twisting change for the Madagascar hissing cockroach using body arching (Fig. S1D, E, i), in the probability of dynamic self-righting for the discoid cockroach using legs (Fig. S3A, v), and in body sliding probability for the American cockroach using wings (Fig. S3C, ii).

## DISCUSSION

Our study quantified self-righting attempts (Fig. 3A), performance (Fig. 3B-D), probability of using kinetic energy (Fig. 5), use of and transitions among strategies (Figs. 4, 6), body rotation (Figs. 7, 8, S1B) and deformation (Fig. S1C-E), and body and appendage behaviors (Figs. S2, S3) in the context of a potential energy landscape (Figs. 2, 7).

### Advantages of dynamic self-righting using rotational kinetic energy

As we hypothesized, self-righting strategies in insects like cockroaches can be dynamic. The ability to self-right dynamically (Fig. 5) by gaining and using pitch and/or roll rotational kinetic energy to overcome potential energy barriers offered the American and discoid cockroaches several performance advantages. First, with all else being equal and confirmed using a physical model (Li et al., 2017), the larger its pitch and/or roll rotational kinetic energy, the faster the body pitched and/or rolled, and the shorter the time to self-right. In addition, although each dynamic attempt costs more energy, as our physical modeling demonstrated (Li et al., 2017), greater body rotational kinetic energy increased the chance of self-righting for each attempt and could save energy overall by reducing the number of failed attempts. Further, pitch and/or roll rotational kinetic energy allowed the animal to reach a broad range of body rotation states of higher potential energy on the landscape (Fig. 7B-E). This gives them the opportunity to overcome energy barriers using a greater number of self-righting strategies. Finally, on slippery surfaces or sand where self-righting using quasi-static leg grasping may be difficult (see Fig. 16B of (Frantsevich, 2004)), pushing appendages rapidly to gain body rotational kinetic energy to self-right can be more effective.

Successful attempts revealed three body and appendage behaviors favoring dynamic self-righting performance (Fig. S2). First, the animal's legs slipped less frequently in successful attempts (Fig. S2A). This was beneficial because leg slipping leads to body yawing, sliding, and premature falling of the CoM, which either dissipates pitch and/or roll rotational kinetic energy or converts it into yaw rotational kinetic energy or horizontal translational kinetic energy that does not contribute to self-righting. Second, the animal's assisting leg(s) more frequently accelerated body rolling and/or pitching in successful attempts (Fig. S2B), adding pitch and/or roll rotational kinetic energy. Third, when using wings to self-right, the animal's body was held during pitching less frequently in successful attempts (Fig. S2C), and therefore did not lose the pitch and/or roll rotational kinetic energy generated by prior wing pushing.

**Body rolling facilitates self-righting by lowering potential energy barrier**

As we hypothesized, for all but one strategy (Fig. 8B), cockroaches rolled their body more during successful than failed attempts as the center of mass rose, because increased rolling lowers the potential energy barrier (Figs. 2D, 7). This is important because ground-based self-righting is a strenuous task. For example, a single hind leg of the discoid cockroach may need to generate ground reaction forces during self-righting as large as eight times that during high speed running (at 8 body length/s) (Full et al., 1995). Using the potential energy landscape model (Fig. 2), if the discoid cockroach self-righted using wings with pure pitching, the mechanical work needed to overcome the highest potential energy barrier (420 μJ) would be seven times that needed per stride during medium speed running (at 5 body length/s) (Kram et al., 1997). Using the observed body rotation during winged self-righting (Fig. 7D, Fig. 8, iv), this mechanical work is reduced by 40% (to 260 μJ). Consistent with this finding, winged self-righting of a cockroach-inspired physical model/robot (Li et al., 2016; Li et al., 2017) demonstrated that body rolling increased the chances of successful self-righting by lowering the potential energy barrier.

Both the American and discoid cockroaches are capable of self-righting using both wings and legs. For both species, using legs with greater body rolling and less pitching is more favorable because it overcomes a lower potential energy barrier than using wings with greater body pitching and less rolling (Fig. 8A-C, ii vs. iii, iv vs. v, red). Given this, the American cockroach's successful self-righting is more stereotyped than the discoid's (Figs. 6, B vs. C; Fig. 7, B, C vs. D, E) partly because its potential energy barrier difference between the strategies is larger. For the American cockroach, the potential energy barrier height difference is 1.7 cm for pitching vs. 0.7 cm for rolling (Fig. 8C, ii vs. iii, red). By contrast for the discoid cockroach, the energy barrier difference was only 1.3 cm for pitching vs. 0.8 cm for rolling (Fig. 8C, iv vs. v, red).

**Advantages of diverse self-righting strategies**

The ability of cockroaches and other insects (Frantsevich, 2004) to use and transition among more than one strategy to self-right offers several possible performance advantages. First, if damaged or lost appendages (Fleming et al., 2007; Jayaram et al., 2011) preclude the use of one strategy, the animal still has an opportunity to self-right using an alternative strategy. Second, the observed unsuccessful strategies such as body twisting and wing flapping (Fig. 4), as well as body yawing and deformation and various body and appendage behaviors (Fig. S1, S3), which seemed not beneficial here, may allow the animal to self-right in novel ways in natural environments by interacting with slopes, uneven and deformable surfaces, or nearby objects (Golubovic et al., 2013; Peng et al., 2015; Sasaki and Nonaka, 2016). Third, even the seemingly stochastic and unpredictable motion over consecutive attempts may be an adaptation to heterogeneous, stochastic natural environments (Kaspari and Weiser, 1999).

More broadly, the use of and transitions among diverse self-righting strategies may be an adaptation for many animals. Studies of ground-based self-righting of beetles (Frantsevich, 2004) and turtles (Ashe, 1970; Domokos and Várkonyi, 2008), and aquatic self-righting of marine invertebrates on underwater substrates (Vosatka, 1970; Young et al., 2006), also observed diverse strategies, including leg pivoting, head bobbing, tail pushing, body dorsiflexion, leg pushing, body flexion, and tail bending.

**Future work**

Our quantification of motion on the potential energy landscape using a simple rigid body only offers initial insights into the mechanical principles of self-righting of small insects. Future work should expand the potential energy landscape by adding degrees of freedoms to better understand how appendage motion and body deformation change energy barriers and stability to result in self-righting (Othayoth et al., 2017). Our quantification of self-righting on a flat, rigid, low-friction surface represents a very challenging scenario. Future experiments should test and model how animals interact with slopes, uneven and deformable surfaces, or nearby objects (Golubovic et al., 2013; Peng et al., 2015; Sasaki and Nonaka, 2016) using potential energy landscapes to reveal principles of self-righting in nature. In addition, given our finding that rolling facilitates self-righting by lowering the potential energy barrier, we speculate that searching to grasp the ground or nearby objects (Frantsevich, 2004; Sasaki and Nonaka, 2016), leg flailing (Othayoth et al., 2017), and body twisting, and during self-righting may induce lateral perturbations to increase rolling. Further, experiments (Rubin et al., 2018) and multi-body dynamics simulations (Xuan et al., 2019) to obtain three-dimensional ground reaction forces of the body and appendages in contact with the substrate will help elucidate the dynamics of self-righting. Finally, electromyography measurements will shed light on how animals control or coordinate (Xuan et al., 2019) their wings, legs, and body deformation to self-right.

**Acknowledgements**

We thank Ratan Othayoth for technical assistance; Ratan Othayoth, Qihan Xuan, Sean Gart, Kaushik Jayaram, Nate Hunt, Tom Libby, Chad Kessens, Peter Várkonyi, and two anonymous reviewers for helpful discussion and suggestions, and Will Roderick, Mel Roderick, Armita Manafzadeh, Jeehyun Kim, and Kristine Cueva for help with preliminary data collection and animal care.

**Competing interests**

The authors declare no competing or financial interests.

**Author contributions**

Conceptualization: C.L., R.J.F.; Methodology: C.L., T.W., H.K.L., R.J.F.; Software: C.L., T.W.; Validation: C.L., R.J.F.; Formal analysis: C.L., T.W.; Investigation: T.W., H.K.L.; Resources: C.L., R.J.F.; Data Curation: C.L., T.W.; Writing - original draft: C.L.; Writing - review & editing: C.L., R.J.F.; Visualization: C.L.; Supervision: C.L., R.J.F.; Project administration: C.L.; Funding acquisition: C.L., T.W., R.J.F.

**Funding**

This work is supported by a Miller Research Fellowship from the Miller Institute for Basic Research in Science, University of California, Berkeley, a Burroughs Wellcome Fund Career Award at the Scientific Interface, and an Army Research Office Young Investigator Award under grant number W911NF-17-1-0346 to C.L., an FSU Jena Academic Exchange Program to T.W., and Army Research Laboratory Micro Autonomous Science and Technology Collaborative Technology Alliances (MAST CTA) to R.J.F.

**Supplementary information**

Supplementary information is available.

## Supplementary Information

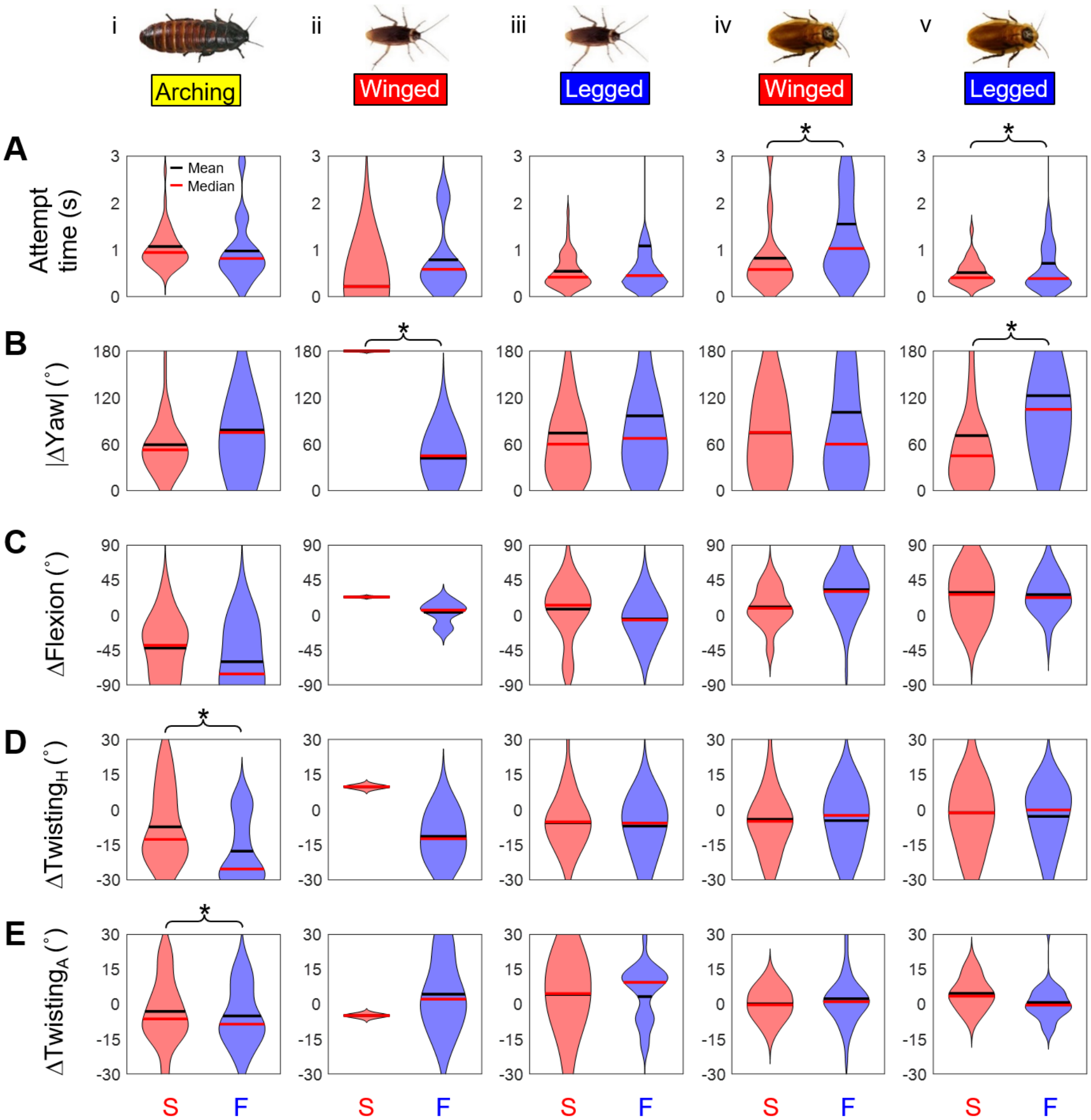

**Fig. S1. Attempt time and body yaw and deformation.** Attempt time (A), body yaw change (B), body flexion change (C), head twisting change (D), and abdomen twisting change (E) when the body was highest for successful (S, red) vs. failed (F, blue) self-righting attempts. (i) Madagascar hissing cockroach using body arching. (ii) American cockroach using the wings. (iii) American cockroach using the legs. (iv) Discoid cockroach using the wings. (v) Discoid cockroach using the legs. We used absolute values of body yaw considering rotational symmetry on the level, flat surface. Data are shown using violin plots. Black and red lines indicate the mean and median. Width of graph indicates the frequency of the data along the *y*-axis. Black asterisks and braces indicate a significant difference between successful and failed attempts ($P < 0.05$, ANOVA). In (C), negative flexion changes in (i) and positive flexion changes in (ii-v) mean increase in hyperextension and flexion, respectively. In (D,E), positive and negative changes in twisting mean increase and reduction in twisting, respectively. See Table 1 for sample size.

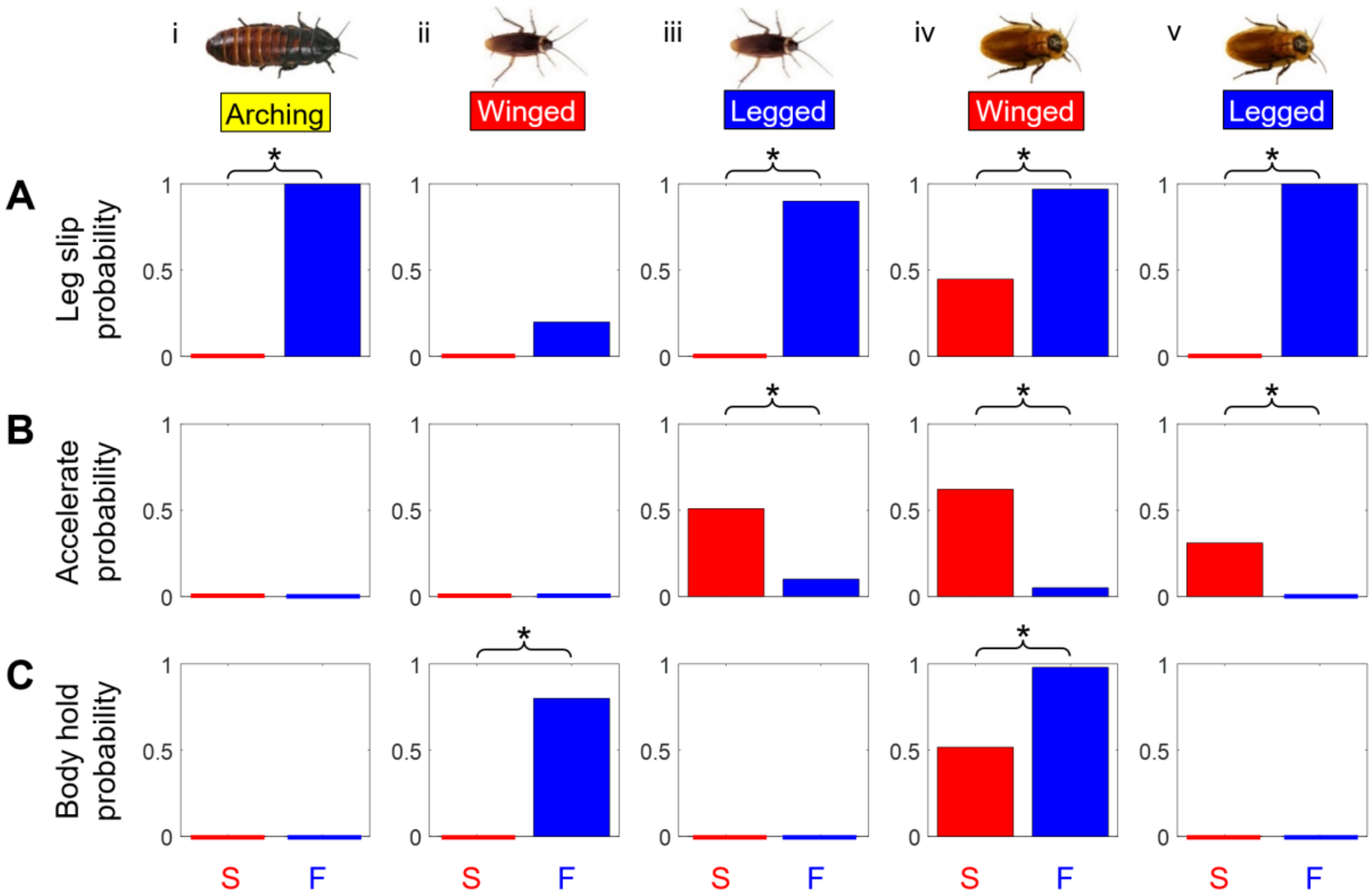

**Fig. S2. Body and appendage behaviors that show a consistent difference between successful (S, red) and failed (F, blue) attempts.** (A) Leg slip probability. (B) Accelerate probability. (C) Body hold probability. (i) Madagascar hissing cockroach using body arching. (ii) American cockroach using the wings. (iii) American cockroach using the legs. (iv) Discoid cockroach using the wings. (v) Discoid cockroach using the legs. Asterisks and braces indicate a significant difference between successful and failed attempts ($P < 0.05$, chi-square test). The large differences between successful and failed attempts in (A, ii-iv) are due to individual variation. See Table 1 for sample size.

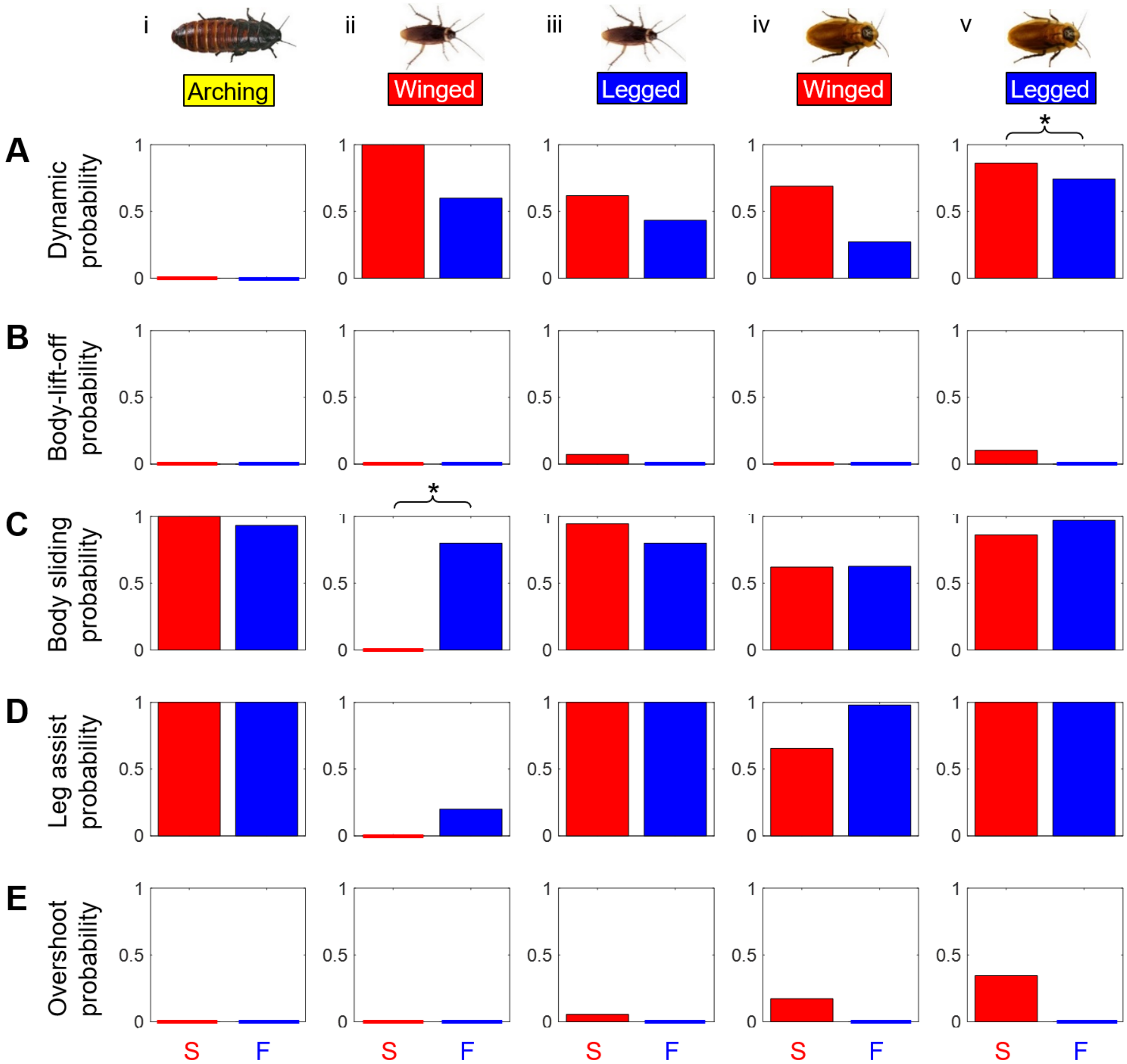

**Fig. S3. Body and appendage behaviors that do not show a consistent difference between successful (S, red) and failed (F, blue) attempts.** (A) Dynamic probability. (B) Body lift-off probability. (C) Body sliding probability. (D) Leg assist probability. (E) Overshoot probability. (i) Madagascar hissing cockroach using body arching. (ii) American cockroach using the wings. (iii) American cockroach using the legs. (iv) Discoid cockroach using the wings. (v) Discoid cockroach using the legs. Asterisk and braces indicate a significant difference between successful and failed attempts ($P < 0.05$, chi-square test). See Table 1 for sample size.

**Movie 1:** Madagascar hissing cockroach self-righting using body arching.
https://www.youtube.com/watch?v=DNJL3ATHbtQ

**Movie 2:** American and discoid cockroaches self-righting using wings.
https://www.youtube.com/watch?v=W3pIJdcv2nw

**Movie 3:** American and discoid cockroaches self-righting using legs.
https://www.youtube.com/watch?v=CoP1-n89DpI